# Coded Single-Tone Signaling for Resource Coordination and Interference Management in Femtocell Networks

*C. Yang, C. Jiang*

*Abstract*—Resource coordination and interference management is the key to achieving the benefits of femtocell networks. Over-the-air signaling is one of the most effective means for distributed dynamic resource coordination and interference management. However, the design of this type of signal is challenging. In this letter, we address the challenges and propose an effective solution, referred to as coded single-tone signaling (STS). The proposed coded STS scheme possesses certain highly desirable properties, such as no dedicated resource requirement (no overhead), no near-and-far effect, no inter-signal interference (no multi-user interference), low peak-to-average power ratio (deep coverage). In addition, the proposed coded STS can fully exploit frequency diversity and provides a means for high quality wideband channel estimation. The coded STS design is demonstrated through a concrete numerical example. Performance of the proposed coded STS is evaluated through simulations.

## I. Introduction

Femtocell networks have been gaining a significant amount of interest for improving coverage and/or increasing capacity by higher spatial reuse via cell splitting in modern wireless communication networks [1]-[3] . Due to the unplanned and dense deployment nature and the restricted access property (a user may be denied for access to the nearest base station and forced to obtain service from a farther base station thereby the receive power from the interferer can be much stronger than that from its serving base station) of femtocell networks, interference can be so severe such that if not intelligently managed the benefits of femtocell networks can hardly be gained [3]-[9] .Tighter coordination among neighboring base stations to ensure QoS and fairness across cells in femtocell networks thus becomes even more crucial than that in the conventional macro network. A straight forward solution is resource partitioning among interfering base stations. This static partitioning makes sense for low bandwidth user-dedicated control channels but is obviously resource-inefficient for data transmission, especially in a dense deployment environment where an overloaded base station can only utilize a small portion of the resources whereas most of the resources could be held up by other idling base stations. Dynamic resource allocation hence makes more sense for data transmission. In this approach, resources are negotiated among interfering base stations or scheduled by a central controller through signaling over the backhaul on an on-demand and maximization of network utility (e.g., fairness, data rates and delay of QoS flows) basis. However, unlike in macrocell networks where base stations are connected via exclusive/leased backhaul, a femtocell is typically connected to the cellular core network via a third party IP backhaul. The delay can be significant and may vary from 10s of msec to 100s of msec. The large and yet unpredictable delay makes signaling through femtocell backhaul unsuitable for dynamic resource coordination and interference management in the case of bursty and delay sensitive applications. Over-the-air signaling is thus a better choice for this type of applications. However, over-the-air signaling presents its own unique challenges. First, over-the-air signaling does not come free. It requires considerable precious wireless resource for the reliable transmission of the coordination message; Second, it requires deep coverage, since the target receivers for the coordination message are the neighboring base stations; Third, due to the broadcast nature of over-the-air signaling, conventional random access signaling [10] suffers from interference among signals sent by different users, and the well-known near-and-far effect [11] . Since the target receivers are the neighboring base stations, the near-and-far effect for this type of application is particularly damaging.

In this letter, we propose a signaling scheme, referred to as coded single-tone signaling, that is particularly efficient for femtocell coordination application. The rest of this letter is organized as follows. Section II gives a detailed description of the coded single-tone signaling scheme. Section III analyzes the proposed coded STS tone detection and decoding. Section IV provides a concrete design example of the coded STS and the numerical results. Section V concludes this letter.

## II. Coded Single-tone Signaling

In the proposed single-tone signaling (STS), a large fraction, if not all, of the energy in an OFDM symbol is transmitted on a single OFDM subcarrier. No energy is transmitted on any other subcarriers of the current OFDM symbol. No information is modulated onto the energized subcarrier (i.e. amplitude and/or phase). It is the location (subcarrier index) of the energized tone that contains information. That is, which subcarrier of this OFDM symbol is energized depends on the content of the message. The message for resource coordination and interference management information is denoted as $m$, which is further represented by $K$ information symbols，$\mathbf{V} = (V_1, V_2, ..., V_K)$, where $0 \le V_k \le S-1$ for $1 \le k \le K$, or more precisely,

$$m(S) = V_K S^{K-1} + V_{K-1} S^{K-2} + ... + V_2 S + V_1 \qquad (1)$$

where $S$ is the number of subcarriers in an OFDM symbol. We hence need $K$ OFDM symbols, with S number of subcarriers per OFDM symbol, to transmit the message $m$, as shown in Fig. 1. The choice of this type of signal has the following advantages: First of all, unlike the most commonly CDMA signals for random access [10][11], the STS does not suffer from the near-and-far effect since all the tones are orthogonal. This property is crucial for this particular application since the target receivers of the signal are the neighboring base stations rather than the serving base station. Second, this STS waveform has low peak-to-average power ratio (PAPR) allowing for a higher power amplifier setting and thereby deeper coverage; Third, since the transmit energy is concentrated on one single subcarrier of an OFDM symbol, the STS tone is much stronger than a regular data tone therefore is easy to be detected even under strong interference environment. If such a strong tone that reaches a base station becomes hard to be detected, this base station will not likely cause significant interference to the sender, i.e., the sender must be out of the interference range of the base station. As a result, STS transmission can be overlaid with other users' data transmission. On the other hand, the interference to the data traffic is also concentrated on a subcarrier of an OFDM symbol. This isolated interference can be most effectively removed by the decoder, thereby causing minimal impact to the data decoding. This means that STS can be transmitted without designated system resource (Therefore, other users don't need to clear the subcarriers for STS). Hence no overhead is incurred. This is in contrast to the conventional random access signaling scheme where a continuous chunk of the total frequency band is set aside for random signaling by multiple users using, for example, PN or Z-C sequences [10]. Fourth, it provides strong wideband pilots/reference signals for accurate channel estimation in a TDD system allowing interference avoidance techniques to be applied for further interference minimization.

However, single tone signaling can also be prone to errors due to fading, other user data interference, and/or multi-user interference among STS signals. Hence it is necessary to encode the STS (non-binary) to obtain certain degrees of error correction capability. Reed-Solomon codes are non-binary codes and achieve the largest possible code minimum distance for any linear code with the same encoder input and output block lengths [12][13]. Hence the Reed-Solomon code is a good fit for the current application. A Reed-Solomon code $(N, K)$ encodes the non-binary information symbols $\mathbf{V} = (V_1, V_2, ..., V_K)$ into a codeword $\mathbf{C} = (C_1, C_2, ..., C_N)$, where $0 \leq C_n \leq S - 1$ for $1 \leq n \leq N$. The error-correction capability is $t = \left\lfloor \dfrac{N-K}{2} \right\rfloor$ and the error-detection capability is $\rho = N - K$, where $\lfloor x \rfloor$ denotes the maximum integer which does not exceed $x$.

In more detail, the parity polynomial for the information symbols $\mathbf{V} = (V_1, V_2, ..., V_K)$ can be obtained through [12]:

$$p(X) = X^{N-K} m(X) \bmod g(X) \tag{2}$$

where $m(X)$ is defined in (1), and $g(X)$ is the generator polynomial which is in the form of :

$$g(X) = (X - \alpha)(X - \alpha^2)...(X - \alpha^{N-K}) \tag{3}$$

$\alpha$ is a primitive element of $GF(S)$ and the coefficients in 错误！未找到引用源。 (2) and (3) are elements of $GF(S)$.

The resulting systematic form codeword polynomial can be written as:

$$c(X) = X^{N-K} m(X) + p(X) = C_1 + C_2 X + ... + C_{N-1} X^{N-2} + C_N X^{N-1} \tag{4}$$

The value of the code symbol (coefficient in the codeword polynomial, $C_n$, $1 \leq n \leq N$) corresponds to the index of the sub-carrier on which energy is transmitted. That is, only one tone is energized per OFDM symbol, and the position of the tone is dependent on the value of the code symbol. Either all or partial of the total energy in an OFDM symbol is transmitted on a single subcarrier depending on the desired coverage range of the coordination scheme. Fig. 2 shows an example of the transmission of the coded STS.

## III. STS DETECTION AND DECODING

### A. STS Tone Detection

STS tone detection is the first step of coded STS detection and decoding. As will be proved in the next sub-section, the coded STS can be designed to be inter-user interference free as long as the coded STS tones are detected error free. Therefore, this sub-section is devoted to the analysis of STS tone detection performance.

The received signal per tone from the $i$th receive antenna can be described as follows:

$$y_i = \begin{cases} \sqrt{E_r} h_i + I, & \text{STS tone present} \\ I, & \text{Otherwise} \end{cases} , \quad 1 \le i \le N_r \tag{5}$$

where $E_r$ is the average received energy of an STS tone which reflects the relative distance between the STS transmitter and receiver ; $h_i$ is the channel gain between the $i$th receive antenna and the transmit antenna and is distributed according to standard complex Gaussian distribution, specifically, $h_i \sim CN(0,1)$ and is independent across $i$; $I$ is the interference (i.e. the traffic data) and can be modeled as a complex Gaussian variable distributed according to $CN(0, \sigma_I^2)$; $N_r$ is the number of receive antennas. Note that $I$ is not a function of antenna index since the interference is the same at all antennas. Multiple antennas thus only provide fading diversity but not interference diversity.

After combining the energy from multiple receive antennas, the detection variable (reflecting the combined energy detected per tone) is:

$$z = \sum_{i=1}^{N_r} |y_i|^2 \tag{6}$$

The cumulative distribution function (CDF) of the detection variable is:

$$P_0(z < x) = F_0(x) = 1 - \exp\left(\frac{-x}{N_r \sigma_I^2}\right) , \quad x \ge 0 \tag{7}$$

when the STS tone is absent, and

$$P_1(z < x) = F_1(x) = 1 - \exp\left(\frac{-x}{\sigma_I^2 + E_r}\right) \sum_{k=0}^{N_r - 1} \frac{1}{k!} \left(\frac{x}{\sigma_I^2 + E_r}\right)^k , \quad x \ge 0 \tag{8}$$

when the STS tone is present.

The false alarm probability (the probability an STS tone is detected on a subcarrier when the STS tone is absent) can be obtained from equation (7) and expressed as:

$$P_{\mathrm{F}} = P_0(z \geq x) = 1 - F_0(x) = \exp\left(\frac{-x}{N_r \sigma_I^2}\right), \ x \geq 0 \tag{9}$$

The miss detection probability of an STS tone can be obtained from equation (8) and expressed as:

$$P_{\mathrm{Miss}} = P_1(z < x) = F_1(x) = 1 - \exp\left(\frac{-x}{\sigma_I^2 + E_r}\right) \sum_{k=0}^{N_r-1} \frac{1}{k!}\left(\frac{x}{\sigma_I^2 + E_r}\right)^k \ , \ x \geq 0 \tag{10}$$

Fig. 3 shows the STS tone miss detection performance in which the false alarm probability is 1%, fading speed is 3 km/h at 2 GHz carrier frequency, and the number of receive antennas is 1,2 or 4. Performance in AWGN channel is plotted as a reference. SIR is defined as the time-domain sample signal to interference ratio. The detection of an STS tone is done by simply looking for a subcarrier with significantly higher energy than its neighbors. From Fig. 3, it can be seen that the STS tone can be detected under low SIR situations. Multiple receive antennas help further reduce the miss detection probability by providing fading diversity. In the next sub-section, we will see that STS tone detection performance directly affects the interference immunity among the coded STS signals.

*B. STS Decoding*

After the detection of STS tones, the receiver obtains a set of code tones on every OFDM symbol with certain errors as well as missed code tones. By applying, for example, maximum likelihood decoding, the receiver finally recovers the original information symbols.

In the presence of multiple users, coded STS signals from different users may overlay on top of each other causing potential interference among different users' coded STS signals. Indeed, $U$ coded STS signals with code rate $(N, K)$ may coexist without causing decoding ambiguity as long as the following inequality

$$K \leq \left\lceil \frac{N}{U} \right\rceil \tag{11}$$

is satisfied. This conclusion can be formally stated by the following proposition:

*Proposition: Assume $U$ $(U \leq S^K)$ distinctive STS signals coded on GF(S) with code rate $(N, K)$ are simultaneously received on the same time and frequency resource. Under perfect tone detection, all $U$ coded STS signals can be decoded to the original information symbols without ambiguity, if $K \leq \left\lceil \dfrac{N}{U} \right\rceil$ is satisfied.*

*Proof:* Consider $U$ $\left(U \leq S^K\right)$ distinctive STS signals coded with rate $(N, K)$ on GF(S) are simultaneously received from $N$ OFDM symbols, free of tone erasures and detection errors. Now randomly select $N$ number of the detected tones, each from one of the $N$ different OFDM symbols. We maintain that

1)    There are at least $\left\lceil \dfrac{N}{U} \right\rceil$ STS tones out of the $N$ selected tones coming from the same coded STS signal among the total number of $U$ coded STS signals. This is the direct outcome from the pigeonhole principle.

2)    For an STS with code rate of $(N, K)$, a minimum number of $K$ STS tones is sufficient to distinguish one coded STS from another. This is sustained by the fact that Reed-Solomon codes are maximum distance separable,

We therefore conclude that if $\left\lceil \dfrac{N}{U} \right\rceil \geq K$, the $U$ STS signals can be uniquely separated from each other without ambiguity. This completes the proof.

For example, a $(14, 2)$ coded STS scheme can allow up to 13 simultaneous coded STS transmissions from different users. The case of particular interest is $K=1$. When $K=1$, (11) holds for any value of $U \leq S$. For example, a (14, 1) coded STS scheme can support arbitrary number of simultaneous coded STS transmissions, i.e., there is no inter-user interference for up to the total number of code words ($S$) of users under perfect STS tone detection. This is a highly desirable multi-user interference resistance property for use in the interference dominant femtocell networks.

## IV. NUMERICAL RESULTS

In this section, we provide a concrete coded STS design example for use in a femtocell network. We will adopt the LTE framework [14] and focus on the downlink for the following discussion. As earlier stated, the coded STS is used for resource coordination and interference management in femtocell networks. A user who is experiencing severe interference from one or more base stations, due to, e.g., entering the coverage of femtocell base stations with restricted access, broadcasts the resource coordination request message (RCRM) via coded STS to all base stations. The base stations that are able to decode the RCRM (therefore, major interferers) can simply clear the resource based on certain criterion or can coordinate power and spatial beams according to the received STS.

The information included in the RCRM in general can be the assigned radio resource identity by the serving base station, traffic priority indicator, the target SINR indicator, etc.

Specifically, the assigned radio resource identity (ID) represents the unique resource. The number of bits of this identity depends on the system bandwidth and the granularity of the resource. For the system with 5 MHz bandwidth ($S$=512 subcarriers), a 2-bit resource ID are typically sufficient to represent four unique sub-bands with 14 OFDM symbols (an LTE sub-frame).

Data traffic priority represents the current traffic priority of the user [15] and should be taken into account by resource coordination. The traffic priority is a metric that is a function of the type of traffic flow (e.g., best effort, delay sensitive QoS flows), packet delay, queue length, and average rate, etc. We allocate 3 bits from the RCRM to represent eight levels of data traffic priorities.

Two bits of the RCRM are used to indicate four levels of the target SINR range of the scheduled data transmission.

It is clear that message collision occurs when two or more mobiles served by different base stations happen to send the messages with the same content, i.e., the same resource ID, the same traffic priority, and the same target SINR. When collision happens, the base station is unable to distinguish the RCRM sent by the users served by the other base stations from its own user. As a result, the base station will not attempt to do interference management for these users on this resource.

To further reduce the collision probability, the base station ID can be programmed into the RCRM. However, base station ID (typically 9 bits) is too long to fit into an RCRM payload. A time-varying hash function can then be used to convert the 9-bit ID into a 2-bit number. The collision probability is hence reduced.

Therefore, the resource coordination request message $m$ in an coded STS consists of 9 information bits, i.e., 2 bits of assigned radio resource ID, 3 bits of traffic priority, 2 bits of target SINR and 2 bits of hashed serving base station identity. With $S = 512$ subcarriers, the 9 bits information $m$ can be represented with $K = 1$ information symbols according to (1) and can be further encoded into a Reed-Solomon codeword with block length $N = 14$ (one LTE sub-frame). In this setup, ideally, we can see that the inequality $K \leq \left\lceil \dfrac{N}{U} \right\rceil$ holds for any $U$ value up to the total number of the code words. This means, in the absence of tone erasure and detection error, up to 512 coded STS signals can be sent simultaneously without interfering with each other.

Fig. 4 shows the decoding erasure and error performance of the proposed coded STS scheme in a multi-user scenario, in which the number of users $U$ is 30, total information bits of signaling is 9 bits, the number of subcarriers in an OFDM symbol is 512, code rate of STS is (14,1) and the number of receive antennas is 1, 2 or 4. Performance in AWGN channel is plotted as a reference. SIR is defined as the ratio of received energy per sample to noise variance in time domain. An erasure is defined as the event in which the base station fails to decode the RCRM sent from a user, while an error is an event in which the base station decodes the RCRM to a wrong but valid message. An erasure causes the base station to fail to respond to the resource coordination request whereas an error causes incorrect response to the request and may result in waste of resource. In Fig. 4, the error rate is controlled below 1%. It is observed that coded STS can operate at very low SIR under multi-user ($U$=30) simultaneous coded STS transmissions. Multiple receive antennas help minimize the fading effect due to the spatial diversity resulting in performance closer to AWGN channel for the four antenna case. In the low SIR region, the up-fades from the frequency selective fading in PedB channel create more opportunities than AWGN channel for the STS tones to be detected, causing less erasure. Note that however multiple antennas does not help reduce interference since the interference (i.e., the traffic data) received is the same at all antennas (remember that coded STS is transmitted on top of the traffic data, i.e., no resource is set aside for coded STS).

## V. CONCLUSIONS

In this letter, we proposed a special signal, i.e., coded STS (single-tone signaling), for over-the-air resource coordination and interference management in femtocell networks. The proposed coded STS scheme has many most desirable properties. First, the coded STS scheme does not require dedicated resource, i.e., coded STS overlays on other users' traffic data and thereby does not incur system overhead. The coded STS does cause interference to other data transmission. However, since, unlike the conventional CDMA signals whose energy are spread over a contiguous bandwidth, coded STS energy is concentrated only on one subcarrier per OFDM symbol the effect on data decoding is minimal. In fact STS tones are easy to be detected, the effect can be further minimized by removing (zeroing-out) the subcarrier where a strong STS tone is detected. Second, the coded STS tones are orthogonal OFDM tones, therefore, has no near-and-far effect. Users transmitting coded STS in the neighboring cells will not be blocked by the local cell users who are transmitting their own coded STS signals, thereby significantly increasing the hearability of the coded STS signals. Third, the coded STS is a single tone waveform with low PAPR, which is especially beneficial for coverage extension and mobile device power amplifiers. Fourth, the proposed coded STS are coded in the way that multiple transmissions of coded STS signals do not create interference among each other. This is another attractive feature for multi-user interference dominated femtocell environment. Finally, coded STS tones spread over the entire frequency band providing maximum frequency diversity for combating frequency selective fading. In the meantime, the wide spread of coded STS tones also produces "free" strong pilots/reference signals across the whole frequency band providing a means for high-quality wideband channel estimation for use in various interference mitigation schemes.

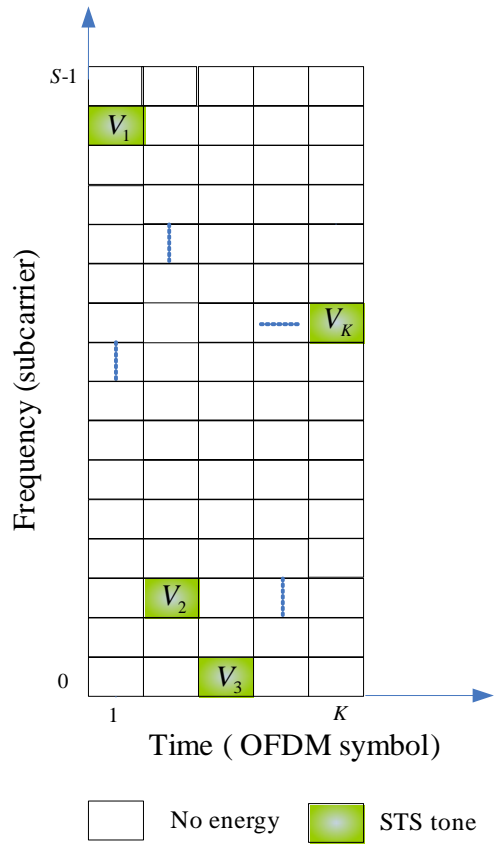

Fig. 1. Illustration of STS.

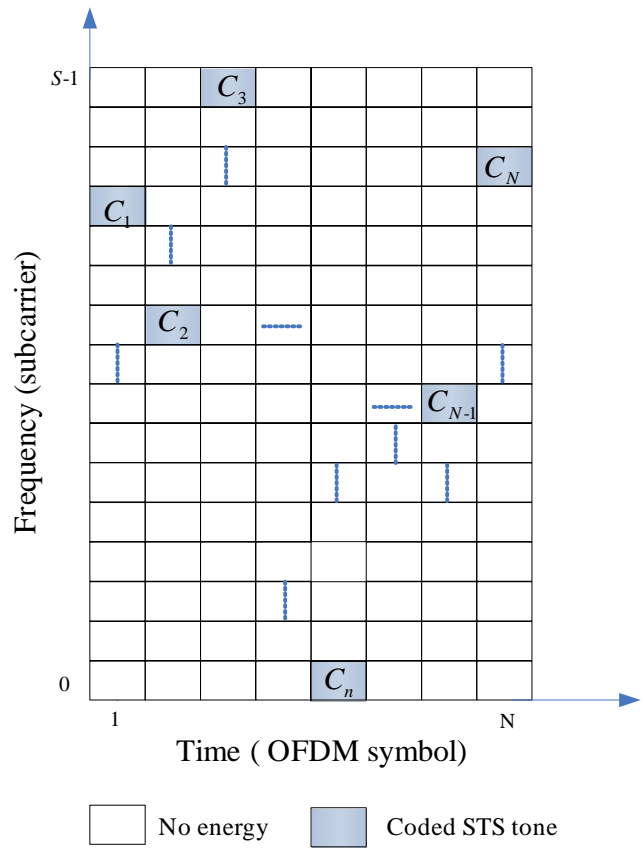

Fig. 2. Illustration of coded STS.

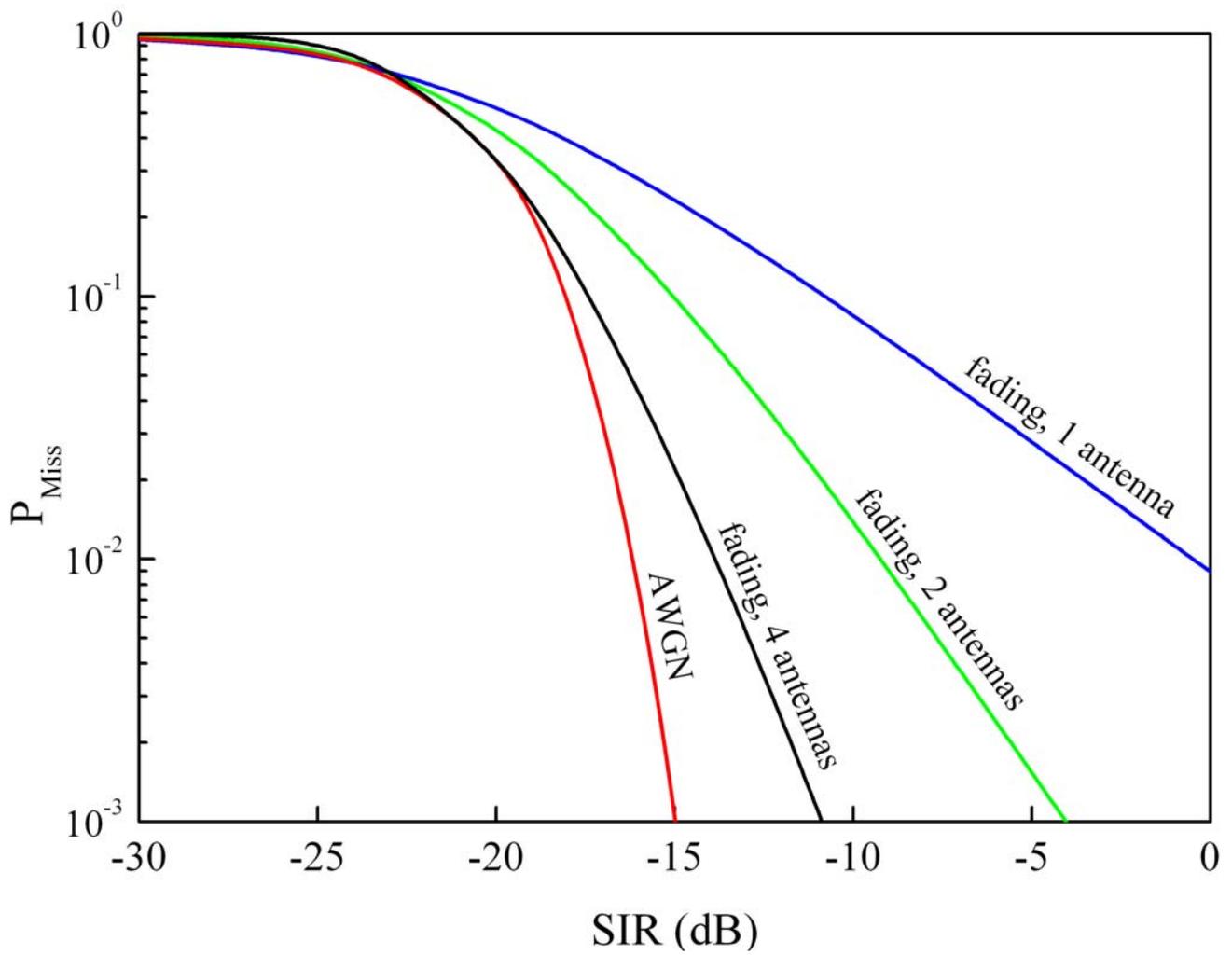

**Fig. 3. STS tone miss detection performance (false alarm probability = 1%; fading speed = 3 km/h at 2 GHz carrier frequency; number of receive antennas =1, 2 or 4)**

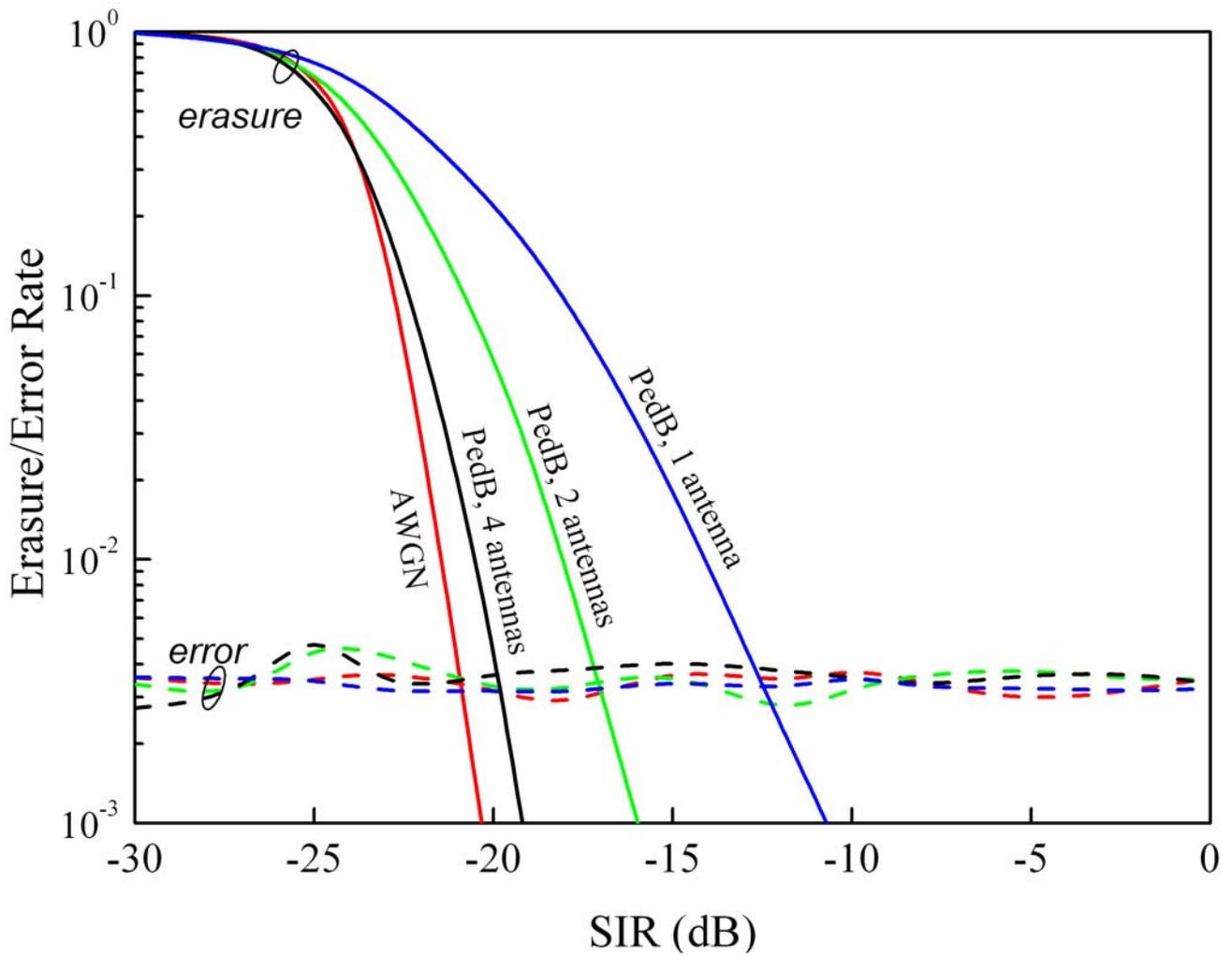

**Fig. 4. The decoding erasure and error performance of the coded STS signaling in a multi-user scenario (30 users; total information bits of signaling = 9 bits; number of subcarriers in an OFDM symbol = 512, code rate of STS =$(14,1)$, fading speed = 3 km/h at 2 GHz carrier frequency; number of receive antennas =1, 2 or 4).**